# Constructing a software requirements specification and design for electronic IT news magazine system

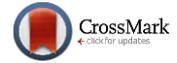

Ra'Fat Al-Msie'deen [1, *], Anas H. Blasi [1], Mohammed A. Alsuwaiket [2]

[1]Department of Computer Information Systems, Mutah University, Karak, Jordan
[2]Department of Computer Science and Engineering Technology, Hafar Batin University, Hafar Al-Batin, Saudi Arabia

A B S T R A C T

Requirements engineering process intends to obtain software services and constraints. This process is essential to meet the customer's needs and expectations. This process includes three main activities in general. These are detecting requirements by interacting with software stakeholders, transferring these requirements into a standard document, and examining that the requirements really define the software that the client needs. Functional requirements are services that the software should deliver to the end-user. In addition, functional requirements describe how the software should respond to specific inputs, and how the software should behave in certain circumstances. This paper aims to develop a software requirements specification document of the electronic IT news magazine system. The electronic magazine provides users to post and view up-to-date IT news. Still, there is a lack in the literature of comprehensive studies about the construction of the electronic magazine software specification and design in conformance with the contemporary software development processes. Moreover, there is a need for a suitable research framework to support the requirements engineering process. The novelty of this paper is the construction of software specification and design of the electronic magazine by following the Al-Msie'deen research framework. All the documents of software requirements specification and design have been constructed to conform to the agile usage-centered design technique and the proposed research framework. A requirements specification and design are suggested and followed for the construction of the electronic magazine software. This study proved that involving users extensively in the process of software requirements specification and design will lead to the creation of dependable and acceptable software systems.

## 1. Introduction

Software engineering is a discipline that is interested in all aspects of software production such as Design, source code, documentation, and so on (Al-Msie'deen, 2019c; 2019a; Al-Msie'deen and Blasi, 2019). Computer software is not only a program but also involves all documentation that is desired by software developers, and users (Sommerville, 2016). A software process is a chain of activities that leads to the construction of a software system. Four basic activities are common to all software processes, which are software specification (i.e., requirements document), development (i.e., design and implementation), validation, and evolution (e.g., requirements change). There are many kinds of a software products, and each demands suitable software engineering methods, tools, and techniques for their development. The basic ideas of software engineering (e.g., requirements engineering and software reuse) are appropriate to all types of a software systems (e.g., stand-alone applications and web applications such as e-commerce applications).

Software developers often access the official documentation (e.g., software specifications) to learn and understand the software system. When the software documentation is absent, it becomes difficult to understand and maintain the software based on its source code only (Al-Msie'deen, 2015; 2018; Al-Msie'deen and Blasi, 2018). Software re-engineering is concerned with re-structuring and re-documenting legacy software systems to make them simpler to comprehend and change. Several studies extract different kinds of documents from software code to help engineers understand the software when its documents are missing. For instance, Al-Msiedeen and Blasi (2021) extracted the software identifiers map from the software code to understand software evolution across the software family. AL-Msie'deen et al. (2014c; 2013b; 2013c) extracted feature models from software variants to understand variability across those variants. While, when the requirements document is available all information about software evolution is included in the preface chapter. Where preface chapter includes a justification for the creation of a new release and a summary of the changes produced in each release.

---

* Corresponding Author.
Email Address: rafatalmsiedeen@mutah.edu.jo (R. Al-Msie'deen)
https://doi.org/10.21833/ijaas.2021.11.014
Corresponding author's ORCID profile:
https://orcid.org/0000-0002-9559-2293

Software processes are the steps involved in creating a software product. Software specification is the process of developing and documenting software requirements documents. The development process is concerned with converting a requirements document into an executable application. Software validation is the process of examining that the software meets its requirements and that it conforms to the real needs of software customers. Software evolution takes place when the software developer modifies current software to meet new requirements.

Software requirements set out what the software should do and describe the constraints on its operation and development. Functional requirements are the services that the software must deliver to the end-user such as the login feature. Non-functional requirements (i.e., emergent properties) specify or constrain features of the whole system such as security. The requirements engineering process consists of requirements elicitation, specification, validation, and management.

Requirements elicitation is an iterative process that consists of requirements discovery, classification and organization, prioritization and negotiation, and documentation (Sommerville, 2016). Requirements specification is the process of documenting the software requirements. The software requirements document is an agreed document of the software requirements. This document is organized in a good manner (such as the IEEE software requirements specification template) so that both system customers and developers (i.e., software stakeholders) can use it. Requirements validation is the process of examining the requirements for consistency, completeness, and realism. Requirements management is the process of managing change to the requirements for a system.

The requirements elicitation stage involves meeting with different software stakeholders (e.g., software architects and customer engineers) to find out information about the suggested software. There are two basic ways to requirements elicitation, which are: Interviewing, where you speak to people about what they do and what they want from the software (i.e., services); and work observation or ethnography, where software engineers watch users doing their work to see what procedures, business documents, and artifacts they use. The software engineer is supposed to use a mix of interviewing and ethnography to obtain the software requirements.

Story and scenario are basically the same documents. They are an explanation of how the software product can be used for some specific job. They describe what end users do, what information they use and generate, and what other software products they may use in this task. The user story is written as narrative text and shows an abstract description of software use (Table 1); the scenario is a structured form of a user story with specific information gathered, such as Task inputs, exceptions, processes, and outputs. A scenario begins with a summary of the interaction. Throughout the elicitation process, specific details are included to produce a complete description of that interaction. A scenario may include a description of the initial state, the normal flow of events, what can go wrong, and the final state. As an example of a scenario, Fig. 1 describes what happens when a professor uploads news to the IMMITN application. Software requirements are usually determined by the customer (i.e., end-user). In this paper, the authors play the roles of the customer and developer.

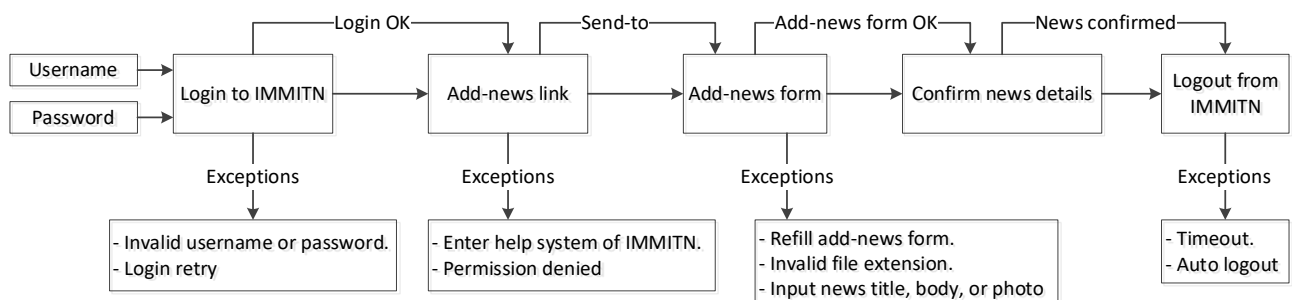

**Fig. 1:** Figure-based scenario for uploading news on the suggested software system

In this paper, we introduce an Interactive Multimedia Magazine for IT News application (IMMITN), which supports user interaction, hyperlink, video, and animation. The application will be implemented through rich interaction and multimedia features. Interactive multimedia technology has been named a hybrid technology. Interactive multimedia combines the storage and recovery abilities of the database with tools for presenting and manipulating database contents. Multimedia has a lot of different meanings, and definitions differ based on the context. For the purposes of this paper, in the context of an interactive multimedia magazine, interactive multimedia is a material that includes some combination of texts, pictures, video, and animation. Interactive multimedia can deliver great amounts of materials in many forms, and deliver them in an application that allows end-users to control the reading and viewing experiment.

Table 1 is an example of a story that we established to understand the requirements of IMMITN. This story describes the IMMITN application. This application uses to deploy up-to-date IT news. You can realize this is a very high-level explanation of the proposed application. Its aim is to facilitate discussion of how the IMMITN application might be used. This high-level description works as a starting point for eliciting the requirement specifications for that application.

**Table 1:** A user story for IMMITN application

| Interactive multimedia magazine for IT news |
|---|
| Ra'Fat is an associate professor at Mutah University (a university in the south of Jordan). He has decided that a software engineering course project should be focused on the deployment of up-to-date IT News. IMMITN is a web-based application. This application consists of four users, which are: guest, student, professor, and admin. The interactive multimedia magazine publishes one volume per year, where the volume consists of two issues. |
| Ra'Fat suggests the IMMITN application, a news-sharing site that allows the admin to check and manage news (i.e., the admin rejects or accepts the news). Students and professors use the IMMITN services to view and add news. When students and professors register to the application and log in, they can immediately use the system to upload news from their mobile devices or personal computers. Finally, the IMMITN application allows guests (*i.e.,* anybody from outside IT college) to browse IT news without any permission. |

The software process is the activities that are related to developing a software product. These days, most systems are being developed using two processes, a plan-driven, and incremental development process. The plan-driven process is a software process where all of the process stages are planned in advance before the software system is developed. This type of software process needs full documentation of software. An example of a plan-driven process is the waterfall model. In a plan-driven process, you plan all of the process stages in advance before starting software development (i.e., requirements and the design are developed separately). The incremental software development process is a process for software development where the software is developed, delivered, tested, and deployed in increments (i.e., requirements and the design are developed together). An example of an incremental process is the agile method of software development (Sommerville, 2016; Boehm and Turner, 2004). This study relies on an incremental software development process, in order to develop the suggested software. Fig. 2 presents the important differences between plan-driven and agile methods to software specification development.

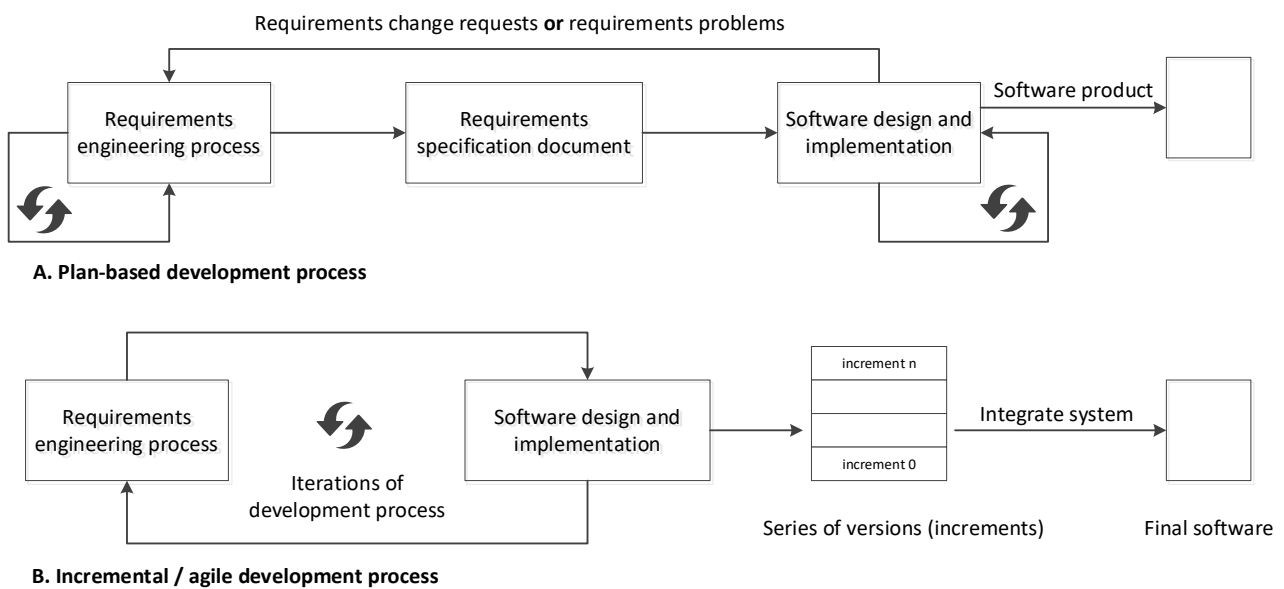

**Fig. 2:** Plan-driven and agile development methods

The agile method is a method of software development that is aimed at quick software delivery. The software product is developed and delivered to customers or customer proxy in increments. In this method, the process documentation is minimized. The focus of this development method is on the software code, rather than software documents.

In this paper, the authors adopted the agile method and user-centered design in order to develop the IMMITN application. The reason for choosing the agile approach is that this approach supports changes that can occur at the level of requirements, unlike the waterfall model. On the other hand, the requirements from the beginning cannot be fully defined. As the software requirements change or as requirements problems are detected, the software design or implementation must be revised, reworked, and tested (Kotonya and Sommerville, 1998). As a software engineer, you should pay attention to the system specifications, as requirements engineering costs about 15% of total system development costs.

Our study is detailed in the rest of this paper as follows. Section 2 presents the related work closest to our contributions. Section 3 outlines the software requirements specification and design development process. Section 4 illustrates the research framework in detail. Finally, Section 5 concludes and presents perspectives for this study.

**2. Literature survey**

This section presents the related works closest to this paper. Al-Msie'deen (2008; 2014) and Alfrijat and Al-Msie'deen (2009) introduced an application called local news WAP/WEB application. Local news application provides users in the rural communities of Malaysia with the appropriate news services that could help them to view and upload numerous news related to their life. The author developed a requirements model for the proposed application. The author suggested a model for the requirements that present the functional and non-functional requirements of the system. Requirement specifications have been written using natural language supported by tables and Unified Modelling Language (UML) diagrams. While the work presented in this paper aims to present the functional requirements of the interactive multimedia magazine application.

Islam and Nofal (2008) suggested replacing the traditional e-book (i.e., training materials for EduWave system in Jordan), with an interactive multimedia book. His work aims to design an interactive multimedia book as training material for the EduWave system and to evaluate this book in terms of interface design and content. The proposed application allows teachers to upload the multimedia contents, then students can view and interact with book pages. His work is similar to the work presented in this research, where multimedia is used to deliver interactive content to users through an interactive multimedia magazine.

In their research, Ayobami and Osman (2013) stated that the reason for the software failure to accept is due to the inability of this system to meet the needs and desires of the clients of the software. The authors proposed a mobile application for fishermen in Malaysia. They evaluated the proposed functional requirements of this application using a qualitative-based users' participatory design methodology. In fact, the application must be validated to ensure that it does what the client wants. In this paper, the authors validate the proposed application based on the predefined software requirements specification.

Choi et al. (2011) suggested an approach in developing e-Textbooks, which distinguished learning from traditional teaching. E-Textbook integrates multimedia and contains more interactive functions than the traditional printed textbook. The same is true for the e-magazine, as it differs from the traditional magazine, as it provides the news interestingly and attracts the reader with the capabilities of multimedia and in an interactive way.

Choi et al. (2014) suggested an interactive e-book reader application, which provides rich features of multimedia such as hyperlinks, video, and animation. Their work proved that HTML-5 based technologies have reasonable merits for an e-book. Their application is very close to the software product suggested in this research, as both applications took advantage of the rich features of multimedia in order to provide the end-user with certain content in an interactive and fun way.

Paz and Boussaidi (2018) offered a software development methodology for an avionics control software system. The authors suggested and followed a requirements specification and design methodology to construct the proposed software system.

Nowadays, the usage of the electronic magazine (e-magazine) is an advanced method of seemly with present generation reading style in obtaining information. This fact becomes clearer with the advancement accomplished in computer technology, the internet, smartphone, and multimedia. Most of the existing studies exploit multimedia in the fields of e-learning, e-book, e-Textbook with limited interactive features. Related work does not provide a study that exploits multimedia features in order to provide information technology news to the user interactively. Compared to the existing works, the e-magazine application of IT news has some features that are common with other applications and has some unique features. The e-magazine is a general IT news application and supports text, multimedia, user interaction, and hyperlink. Most existing studies are designed to obtain the software requirements specification to construct the software system in a systematic way. The originality of this study is the construction of software specification and design of the electronic magazine by following the Al-Msie'deen research framework which exploits the agile usage-centered design technique.

**3. Software requirements specification and design development**

This section presents the agile usage-centered design technique and functional requirements of IMMITN software. It also introduces the Al-Msie'deen research framework.

### 3.1. Agile usage-cantered design technique

In order to propose a professional software requirements document, for the proposed software, authors based themselves on the user-centered design and agile techniques. Agile and user-centered design are two significant development techniques for confirming that a software application has the best user experience (Jurca et al., 2014). The user-centered design technique is a design technique that emphasizes user study, user interface design, usability testing, and assessment (Caballero et al., 2016; Patton, 2002). The agile technique is a software development technique that focuses on satisfying customer needs and changing customer requirements, even late in development (Sommerville, 2016). Both techniques aim to improve the user or customer viewpoint on product quality, they agree on the importance of user or customer contribution in interactive software development processes, and both techniques highlight the importance of developing team cohesion. The agile technique focuses on the customer and the key principles of user-centered design are an early and continuous emphasis on software product users.

One goal of this study is to achieve a user-friendly interface for IMMITN software and to meet users' needs. The user acceptance of any software lies mainly on how it meets his or her behavioral intention, and this is mostly identified by the degree to which the software meets the wanted needs of the users (Ayobami and Osman, 2013; Davis, 1985). Shneiderman et al. (2016) suggested that in a user-centered design technique, user behavior research should be undertaken by studying his or her working environment (i.e., workplace and task list). Authors use the agile usage-centered design method (Valacich and George, 2019; Patton, 2002; 2004; Constantine and Lockwood, 2002) in order to develop the suggested application. This method is the result of combining both agile and user-centered design techniques. Frequent user involvement in software development is a perfect method to ensure that requirements are captured precisely and directly implemented in software design. However, such continuous interaction works great when the software development team is small, as is the case in the IMMITN application. As well, it is not always possible to continuously reach users throughout the project development period. Therefore, agile software engineers have come up with other ways for well including software users in the requirements elicitation process. One such way is named agile usage-centered design. In this work, functional requirements captured from software users and engineers are captured as prototype screens.

### 3.2. Software requirements specification or SRS

After the software requirements are discovered, these requirements are documented by writing them in the requirements document (i.e., software requirements specification). The process of writing the software requirements is called the requirements specification process. Ideally, the software requirements must be complete and accurate (Sommerville, 2016). Software stakeholders understand the requirements in different ways, so the requirements must be written in an accurate way so that they are clear to all readers and do not allow multiple interpretations. In general, software requirements are read more than they are written. Thus, a software engineer should invest time to write clear and comprehensible requirements documents. User requirements of any software describe the services provided to the user in high abstract. On the other hand, the system requirements describe the user requirements in more detail. Both the user and system requirements form the functional requirements of any software. The agile technique integrates the requirements engineering process with software design and implementation. In addition, the agile method develops software as a series of increments. Where most agile methods propose that software products should be constructed and delivered incrementally.

Software requirement is typically written on the SRS document as a paragraph of natural language text enhanced with diagrams. Different types of requirements are wanted to transfer information about software to different kinds of stakeholders. Table 2 displays the difference between user and system requirements. This instance, from IMMITN software, shows how a user requirement may be extended into numerous system requirements. The reader can see from Table 2 that the user requirement is general and abstract, while system requirements offer more details about the service of the software that is to be implemented.

Requirements document is important when software is outsourced for development, when different teams construct different fragments of the software product, and when a complete analysis of the requirements is required. Since the requirements change very quickly, the requirements document becomes out of date as soon as it is written down, so the effort of the engineer (i.e., document writer) is mostly wasted. Therefore, a full document is not required for agile and user-centered design techniques. Instead of a formal SRS document, the agile technique commonly gathers software requirements incrementally and writes these on the SRS document as short paragraphs. Then, the user prioritizes software requirements for implementation in the following increment of the software product (Sommerville, 2016; Kotonya and Sommerville, 1998). Developing the full software requirements document is a costly and time-consuming process. Thus, the software requirements document must define the minimal functional requirements for the software system.

Functional requirements are written in natural language enhanced by suitable diagrams and tables in the SRS document. A data flow diagram (DFD) is a depiction of the movement of the main information flows among external entities, processes, and information stores in the software (Valacich and George, 2019). The highest level of the DFD is termed a context diagram. The context diagram consists of one process, data flows, external entities (*i.e.,* sources/sinks), and no data stores. The single process, labeled with *0*, represents the whole software (Fig. 3).

Table 2: An example of user and system requirements of IMMITN

| User requirement definition |
|---|
| 1. The IMMITN software shall allow the student and professor to register as a new member of a software system |
| System requirement specification |
| 1.1 In order to add IT news to IMMITN software, the student and professor shall click the registration button |
| 1.2 The system shall respond to his or her request by showing the registration form |
| 1.3 The system shall allow new users to fill all registration form fields in order |
| 1.4 The system shall allow students or professors to press the registration button after the user filled in the registration form |

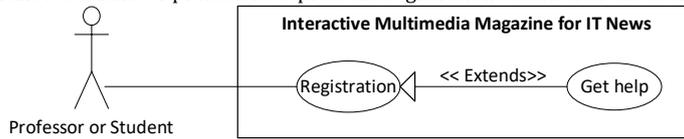

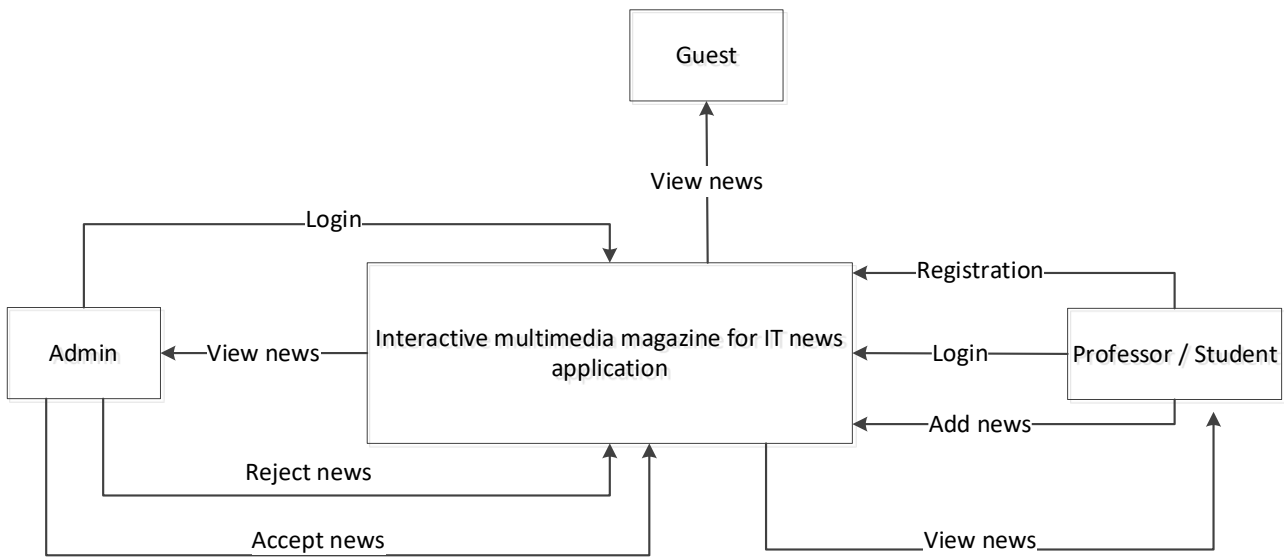

**Fig. 3:** The context diagram of IMMITN software

The level-0 DFD diagram shows system main processes, flows of data, and stores of data at a high level of abstraction. The external entities of level-0 are the same in the context diagram. The level-0 DFD diagram gives the main processes in the software at the highest level. Each process in the level-0 DFD diagram has a number that ends with a zero (matching the level number of the diagram).

UML is a graphical notation applied to object-oriented (OO) software design that consists of numerous kinds of software models that offer different views of software aspects. The use-case diagram is a UML notation kind that is exploited to detect use-cases (i.e., processes) and graphically show the users implicated in a software system (Al-Msie'deen et al., 2014a; 2014d). The use case diagram of the IMMITN application consists of four actors, and seven use cases. It is not enough just to show the use case diagram in UML notation. A detailed explanation must be provided for each use-case. A detailed tabular presentation of each use-case is presented by use-case description. The use case is a description of one kind of interaction between a user and a software system. The use case diagram must be enhanced with extra information to fully describe its use-cases. Thus, the use case description assists software designers to recognize objects and processes in the software system. This gives designers a comprehension of what the software is planned to make (Sommerville, 2016). The use case description consists of a graphical notation and structured natural language text (Table 3).

The sequence diagram shows the sequence of interactions between users and the software and between software objects or screens (Sommerville, 2016; Valacich and George, 2019). The sequence diagram displays what inputs are needed and what outputs are produced for one type of interaction between user and software system. The reader of the sequence diagram follows the sequence of interactions from top to bottom. The users and objects are listed on the top of a sequence diagram. For each object and user in the diagram, there is a dotted line drawn vertically from it. To illustrate the interaction between two objects in this diagram, there is an annotated arrow with details. Also, on the dotted lines, there is a rectangle showing the lifeline of the object (Fig. 4).

**Table 3:** The use case description of login use-case

| | |
|---|---|
| Use-case name | Login |
| Use-case notation | 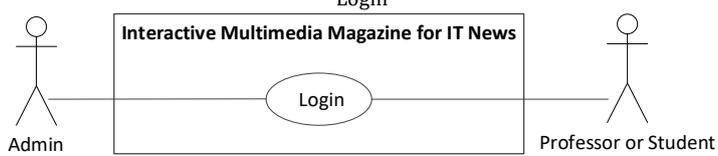 |
| Actor(s) | Administrator, student, and professor |
| Brief Description | The login use-case allows system users to access their main pages |
| Pre-Conditions | The software user must log in to his or her account by entering username and password |
| Basic Flow | 1. The software user needs to insert his or her username and password into the login form<br>2. The software user needs to confirm the login process by pressing the login button<br>3. The software will respond to his or her order by verifying the login information after checking the software database<br>4. The software will send the user to his or her main page |
| Exceptional Flow | Wrong username or password / Refill the login form |
| Post-Conditions | The software user can access his or her main page |

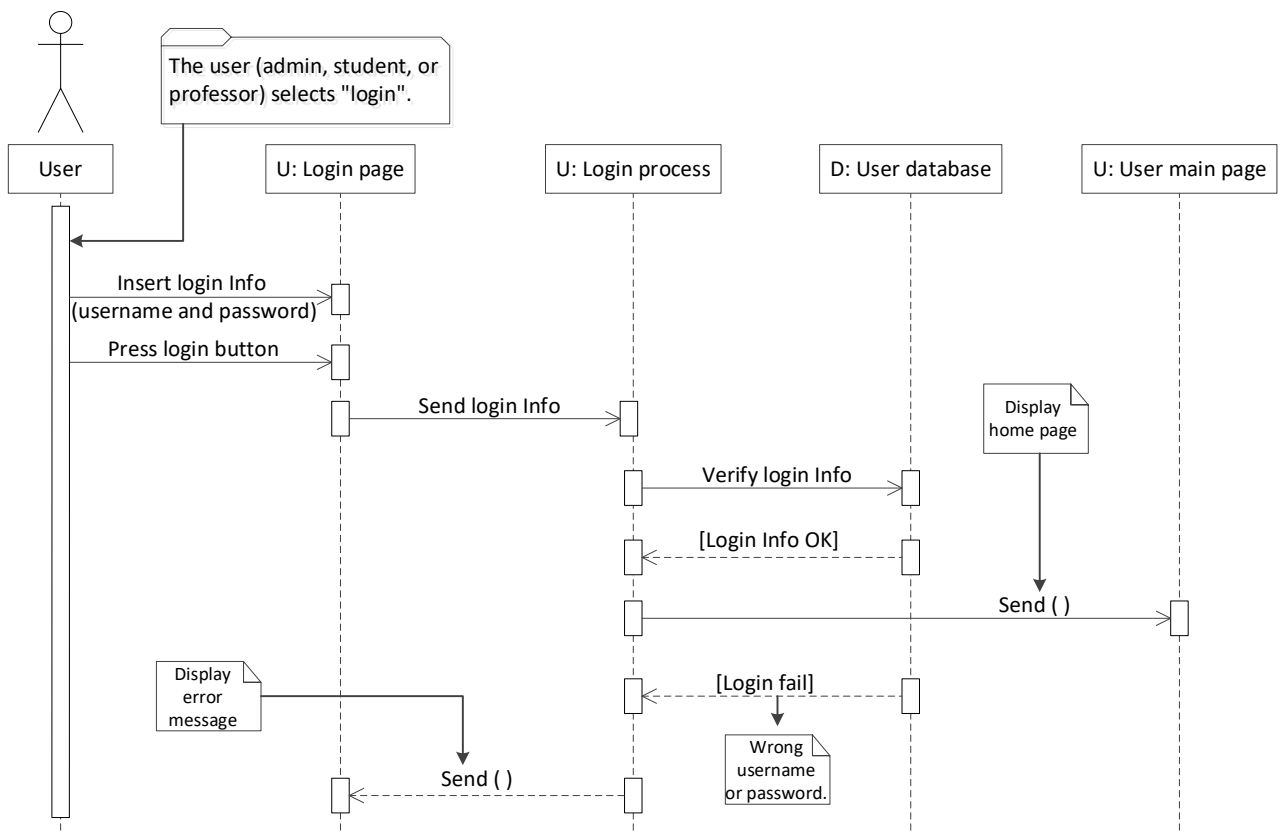

**Fig. 4:** Sequence diagram for the login process

The architectural models of a software product can be exploited to concentrate discussion around the software specifications or design. Thus, software architecture plays an important role in the identification of software requirements specification (Booch, 2005; Bosch, 2000). The software architecture document is a design document and aims to document the software design. Software architecture is used to stimulate discussion about system requirements at an early stage of the design process. The outcome of the software architecture design process is an architectural model that defines how the software is organized as a collection of cooperating components. The architectural model displays these components and the relations between them (Sommerville, 2016).

The architecture of a software system may be documented from different views (i.e., logical, development, process, or physical) (Kruchten, 1995). For example, a software architect can document the software's architecture from a conceptual view (Hofmeister et al., 2000). The software architectural view is an explanation of software architecture from a specific view. It is impossible to give all information around a software architecture in one diagram, as an architectural model can just display one view of the software. The conceptual view is an abstract view of the software that can be the base for decomposing user requirements into detailed software specifications. Conceptual view aids software developers to make choices about components that can be reused in similar products (i.e., software product line) (Al-Msie'deen et al., 2014b; 2013a; Eyal-Salman et al., 2012).

In this paper, the conceptual view of software is developed in the software design process. Software engineer uses this view to explain the software architecture to different stakeholders. The main goal of using the conceptual architecture view in this study is to identify the main components of the proposed software and to capture the main relationships between those components.

The proposed software provides the user with the ability to browse the news through the e-magazine. There are some features that should be considered in the design of the e-magazine. These features are intended to make browsing the news easier and more interactive. The e-magazine design should meet at least the following features: Zoom in and zoom out, search, auto flip, stop auto flip, share, first and last page, previous and next page, print, full screen, select text, page number, and page flip sound.

Non-functional requirements are software requirements that are not directly interested in specific features delivered by the software to its end users. These requirements typically specify or constrain the properties of software as an entire. They may link to emergent properties of a software system such as reliability, and availability. Otherwise, they may describe constraints on the software implementation (e.g., abilities of I/O devices, data representations). Table 4 is a description of some non-functional requirements of the IMMITN system.

**Table 4:** Description of some non-functional requirements of IMMITN

| Requirements | Description |
| --- | --- |
| Performance | *Response time*: IMMITN system should be able to retrieve user requests within 5 s. |
| | *Scalability*: IMMITN system should be able of supporting no less than 100 users at a time when applied. |
| | *Platform*: IMMITN system should be capable to operate in Google Chrome and Microsoft Edge. |
| Security | IMMITN system must guarantee that data about diverse types of user transactions must be treated in a protected channel. |
| Usability | Users with diverse background knowledge can simply deal with the system. |
| Support | Network, application, database, and administrative support should be provided 24/7. |
| Availability | IMMITN system should be capable to deliver services when requested 24/7. |
| Safety | To avoid possible data damages or losses, the IMMITN system must have a data recovery method. |
| Reliability | IMMITN system should be capable to deliver services as specified. |

Functional and non-functional requirements should be included in the SRS document. This paper focuses on the Functional requirements of the proposed system.

### 3.3. The research framework

Today's applications attempt to resolve complex and real-world problems. A strict research methodological framework can ease the encumbrance of this complexity. Without such a framework, software engineers can find it difficult to meet product functional requirements. Moreover, as software requirements change and grow, software that does not have a basic architecture and design will have trouble adapting to these growing requirements. This subsection introduces the suggested research framework.

The user-centered design and agile techniques focus on user participation in the software development process. Software developer aims to realize a user-centric application and meets the user functionalities or expected features of the product. In this work, the authors assume that the user or customer is responsible for defining the requirements for the required program, and at the same time it is he who evaluates the program and makes sure that the program performs the required functions from it.

A software engineer can use many methods to determine a system's requirements, and he or she can use a combination of these methods to determine a system's requirements. There are some methods that are called the traditional methods to determine the system requirements. Examples of these traditional methods are interviewing, observing users at the workplace, and analyzing procedures and supplementary documents such as business documents (i.e., forms and reports). On the other hand, there are some contemporary methods for determining the requirements of a system such as joint application design, computer-aided software engineering tools, and prototypes (Valacich and George, 2019). In this study, the authors determine the functional requirements of the system and evaluate them. The authors use the agile usage-centered design method to develop the proposed system; and from here the system requirements are determined through the continual user involvement throughout the entire system analysis, design, and validation process.

In this study, a content analysis of online materials was not performed due to the lack of applications similar to the software proposed in this research. If there are previous studies similar to the proposed software, they can be analyzed and used to form the functional requirements for it. Fig. 5 shows the research framework based on the authors' design and thought (called the Al-Msie'deen research framework).

In this framework, the problem to be solved is initially defined. Then, user and software requirements are fully understood and translated by designing a graphical user interface (GUI). After that, the user validates the software screens that reflect the functional requirements of the software to be developed. After that, the requirements are documented and written in the software requirements specification. After that, the program is developed in increments, and delivered incrementally to the user for validation (i.e., verifying that software meets the requests and expectations of the user) and verification (i.e., verifying that software meets its requirements) (Sommerville, 2016). The research framework supports the continual user involvement in the design and implementation processes in addition to requirements validation.

## 4. Research framework step-by-step

This section presents the research framework stages in detail. Also, it provides an overview of the software development process by following the proposed research framework. It also displays the design of the IMMITN software system.

### 4.1. Problem statement definition

A problem statement definition is a brief description of an issue to be addressed. The problem statement determines the gap between the current and desired situation. Thus, the problem statement in this paper is identifying the functional requirements of the IMMITN software to be used through different types of software stakeholders. The main outcome of this research is the software requirements specification document which contains functional requirements and software design.

### 4.2. Functional requirements and software design

The heart of this study is to define software requirements and translate them into a design that meets these requirements. In the research framework proposed in this study, after defining the problem to be solved, the user and software requirements are determined and accurately understood. Then, the software designer translates these requirements into a graphical design for the software. After that, software requirements and design are documented in the initial requirements document. Next, the requirements are validated and documented via an SRS document. Then, the coder implements the software. The following subsections explain in detail these stages.

#### 4.2.1. Understanding user and software requirements

The primary condition of resolving a problem is the comprehension of the problem. The software engineer must fully understand the problem to be solved. In order to produce software that suits the user's desires, the user must be involved in the requirements engineering process. It is necessary to understand the user and software requirements in order to obtain the desired software. User requirements are usually used in drafting the contract between the customer and the developer of the software. User requirements state the services that the customer wants to obtain from the software system. Sometimes the customer may be unsure of the services they want from the software. In this case, the developer can assist the customer in proposing software requirements through the requirements proposal document.

Understanding the requirements of the user is indispensable for understanding the requirements of the software system. Software requirements are the same as the user requirements with more details. Consequently, the software requirements are the services that the software provides in detail. A thorough understanding of the problem helps both the user and the developer to do their work professionally. The software cannot be designed if the services it provides are incomprehensible and unclear. Thus, understanding the problem is half of the solution.

Usually, the customer proxy determines the user requirements. For example, the customer requests the developer to add a login service to the software to allow him to use the main services of the software. Login service is described as Access to the software services is done through the login form, where the user enters the corporation's email address as a username and his employee's number as a password. After that, he presses the login button to be transferred to his main screen. Software engineers and customers work together on software requirements. Where the requirement is described in detail. The developer and the user must understand the two types of requirements accurately.

In the case of software systems similar to the software to be developed, an analytical study can be conducted. Content analysis of these similar studies helps reuse some of the requirements. Reuse of requirements saves the cost of producing the software. In the case of the absence of similar studies, the software engineer develops the requirements from scratch. Within the framework proposed in this study, the user is involved in the software development process as a full-time team member. Consequently, there is rarely a misunderstanding in determining the software's functional requirements.

#### 4.2.2. Graphical user interface design

The graphical user interface design is the process of designing the method in which software users can access software services. Also, it is the method that information generated by the software is shown (Sommerville, 2016). The graphical user interface design aims to translate user and system requirements into concrete visual interfaces. In this study, the software requirements are translated into a user interface through the use of prototypes. IMMITN software system consists of two parts (i.e., news management and e-magazine). The four users of the system (i.e., admin, professor, student, and guest) can read the news through e-magazine archives without any permission. While news management allows three users (i.e., admin, professor, and student) to deal with it. Fig. 6 shows the GUI for the registration form from IMMITN software.

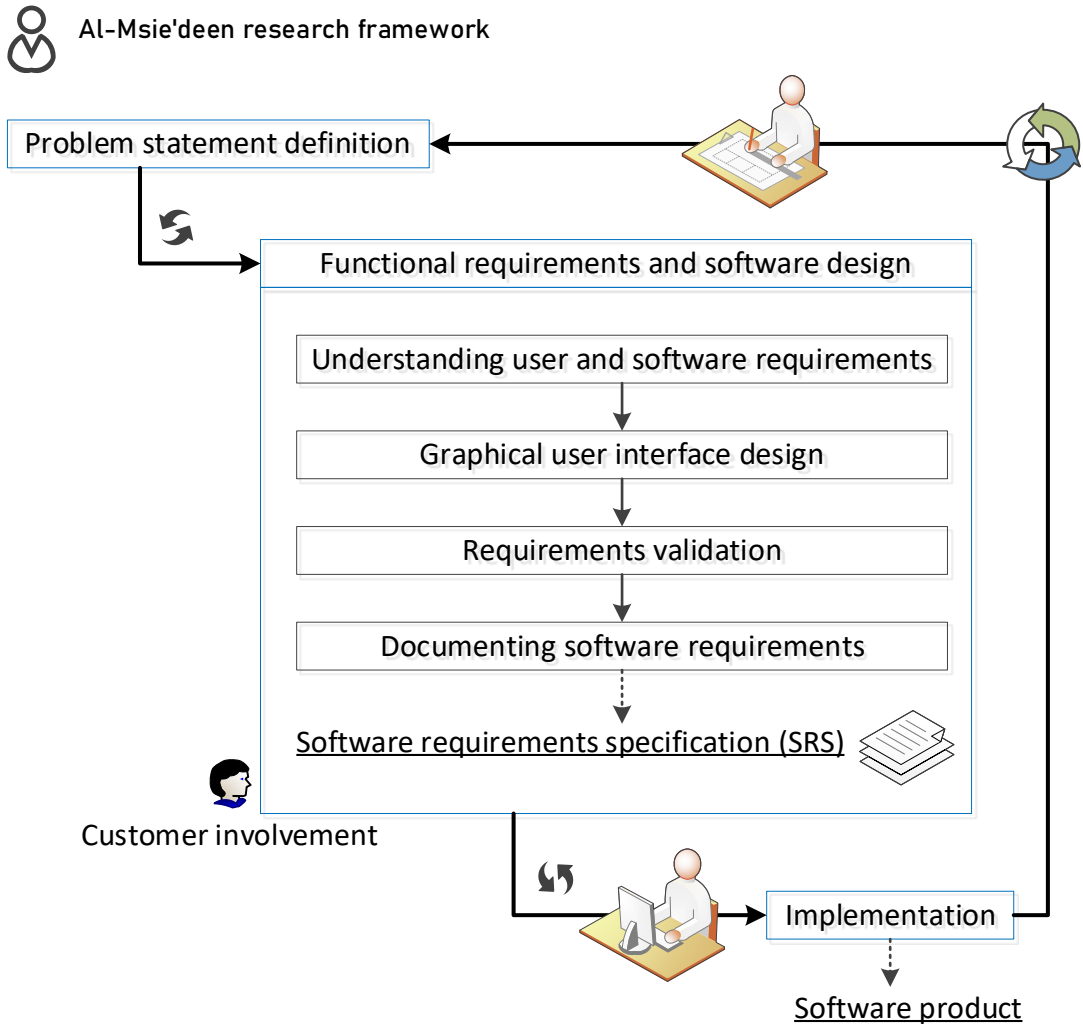

**Fig. 5:** The research framework

**Fig. 6:** Design of the registration form from IMMITN software

Usually, GUI contains forms and reports. Form considers as a business document. It holds some predefined data. Also, it contains many parts where extra data are to be filled in Valacich and George (2019). The process of forms design is a user-centered activity. Thus, forms design activity includes an understanding of the intended user, their responsibilities and goals, data needs, experience levels (i.e., novice, expert, or an advanced user). Fig. 7 presents the design of adding news form for IMMITN software.

**Fig. 7:** Design of add news form for IMMITN software

Human-computer GUI explains the methods in which software users interact with a software system. A software engineer can use several methods to design software GUI. For example, the designer can use the paper prototype to design the interfaces of any software system. As another example, software designers can use the NetBeans IDE GUI builder to design software interfaces. The NetBeans IDE has several exciting features aimed at visual web application design. An effective software designer must comprehend the existing resources and how to use them. Also, the software designer must become a professional user of the GUI environment. The designer must have good communication skills because his work requires him to communicate and interact with the software's users.

Two software users play the same role in adding news to the system, namely, the professor and the student. In order to add the news, they must first register with the system, then log in with their username and password. After logging into the software, they add the news through the add news form, and then they submit the news to the software administrator. The system administrator logs in with the username and password given to him by the dean of the faculty of information technology. After logging, he or she manages the news, by viewing the content of the news added by the professor or student. Then, the software administrator accepts or rejects the news according to the value of the news and its content. If the news is accepted, it will be linked to the appropriate issue of the magazine's volume.

### 4.2.3. Requirements validation

In requirements engineering, the requirements validation process is a requirements analysis activity that aims to discover some potential problems with these requirements. At the stage of validation of requirements, usually, a requirement engineer will discover some problems with software requirements. In this case, the stockholders will negotiate about these problems in order to find a compromise and get an agreed upon requirement. Thus, requirements validation is the procedure of checking that requirements describe the software that the user actually needs (Sommerville, 2016).

After the requirements are elicited, analyzed, and written in the initial requirements document, these requirements are checked to discover some problems between the requirements such as requirements conflict. In the event of problems in the requirements, they are corrected by consulting the stakeholders and then writing them in the final form in the SRS document. Requirements validation is critical because errors in the requirements document can lead to a significant rework budget when these errors are revealed through software development or after operating the software product. The requirements in the requirements document must be checked to ensure that they meet the user's real needs. Also, the requirements in the requirements document must not conflict. On the other hand, requirements must be checked to confirm that they can be implemented through the proposed software budget and schedule. Requirements reviews and prototyping are examples of some requirements validation techniques that can be used. In this study, by following the proposed research framework, the requirements are validated in the requirements document by engaging the user in previewing and approving prototypes. In the case that some problems are discovered, a discussion takes place with the user, and changes are made to the prototype until reaching the agreed requirements.

### 4.2.4. Documenting software requirements

Requirements engineering aims to discover, analyze, validate, and document software requirements. After issuing the initial version of the software requirements (initial requirements document), all the requirements in the requirements document are validated. After agreeing on the final requirements of the software, it is documented in the final form of software requirements (software requirements specification). Thus, software requirements specification is the procedure of officially documenting the user and software requirements and producing a software requirements document.

A software requirements specification is a document that describes the nature of software. In other words, an SRS document is a manual of software. It is prepared before the programmer codes the software. In our work, this document or report is produced in minimal detail based on an agile approach for software production. SRS document contains the functional requirements of the suggested software in a clear and precise manner. The functional requirements of software were written completely, precisely, unambiguously, and without any contradictions. Different software stakeholders must read the document without any problems. This document includes software design documents such as an architectural model of software and UML models of software. Thus, SRS included the agreed and documented software features or services.

SRS consists of a collection of chapters. For instance, the preface chapter includes information about the context of software and its release. Then, the functional requirements chapter includes all functional requirements definitions. In addition, the SRS report consists of a table of contents, indexes such as a list of figures and tables, a glossary that defines the main terms, abbreviations, and acronyms, and appendixes. This document is very useful and important for software engineers. For example, in order to reuse software requirements, it is important to be documented. For maintenance engineers, they are not able to maintain the software without understanding its services or functionalities.

Software documents should work as a communication channel between various software stakeholders. In addition, they should act as an information store in order to be used by different stakeholders (e.g., software coders). Software documents should offer information for software managers to assist them during software development processes (e.g., schedule and budget). On the other hand, user documents (or software manuals) should tell end-users how to use or manage the software in their workplace. End-users employ the software to perform work tasks, whereas software admin is responsible for controlling the software used by end-users (Barker, 2002). Fig. 8 shows a well-structured software tutorial for IMMITN software.

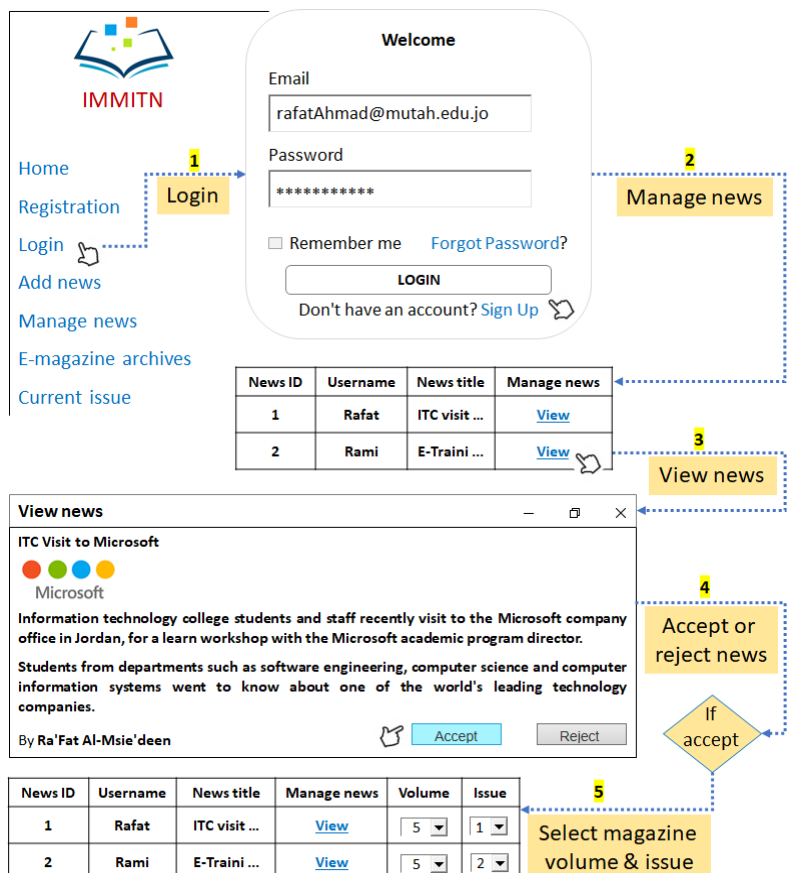

**Fig. 8:** A well-structured software tutorial for IMMITN software

The software manual in Fig. 8 explains the role the system administrator plays in this software. As the arrows indicate the steps followed by the admin to control the news added by the end-user.

### 4.3. Implementation

This work focuses on defining software requirements and designing the software based on the services required by the user or customer. Therefore, this work does not provide a full implementation of the software, though parts of this work have been implemented. The e-magazine includes text, images, and video clips. Also, e-magazine supports user interaction and animation. Fig. 9 shows two pages of the e-magazine, which support text and image.

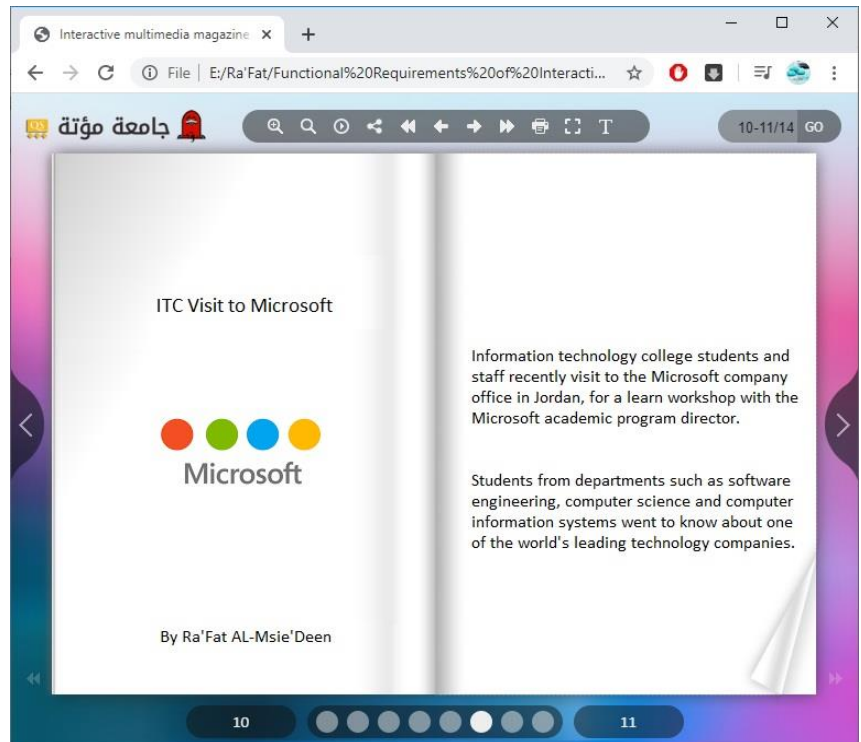

**Fig. 9:** E-magazine of IMMITN software, with an example of IT news

IMMITN software has been implemented as a web-based application. Authors used HTML and CSS for magazine page layout. Also, the authors utilized JavaScript and PHP for scripting and user interaction with the e-magazine. HTML has reasonable advantages of e-magazine application such as ease of implementation, rich features for user interaction and multimedia, and the majority of software engineers are familiar with these technologies. All pictures and tables used in this study are available on the author's website.

### 5. Conclusion

This paper proposed an original approach to develop software requirements document based on the Al-Msie'deen research framework. This research framework was implemented on the IMMITN software system. Results demonstrated that involving users extensively in the process of defining software requirements and design leads to achieving a dependable and acceptable software system. The study also showed the importance of the requirements engineering process in software design. Results also showed that combining agile and user-centered design techniques lead to frequent user involvement in software development. Where user involvement perfect way to ensure that requirements are captured precisely and directly implemented in software design. Regarding future work, the authors plan to add some features to the e-magazine such as thumbnails. The authors also intend to create an e-magazine online manual (user and reader help) for IMMITN. Furthermore, the researchers plan to implement the software based on the requirements and design determined in this research.

Also, the authors plan to apply the idea of the interactive multimedia magazine in institutes of higher education in Jordan by using interactive multimedia books in order to improve the process of distance learning (Alsuwaiket et al., 2019).

Finally, the authors plan to apply the tag cloud visualization technique (Al-Msie'deen, 2019b) on the SRS document to show the most frequent words across this document.

**Compliance with ethical standards**

**Conflict of interest**

The author(s) declared no potential conflicts of interest with respect to the research, authorship, and/or publication of this article.